COMUNICACIONES

# Estudio exploratorio sobre las competencias informacionales de los estudiantes de la Universidad de La Habana[1]

Carlos Luis González Valiente
Yilianne Sánchez Rodríguez
Yazmín Lezcano Pérez

**Resumen**

La presente indagación expone los resultados de una encuesta relacionada para la identificación de competencias informacionales, aplicada a estudiantes de la Universidad de La Habana, como resultado de prácticas investigativas de estudiantes de Ciencias de la Información de dicho centro. Se utilizan varios métodos como la encuesta, la entrevista no estructurada a expertos y el análisis documental y de contenido. El cuestionario fue elaborado sobre la base de tres variables básicas: búsqueda de información, análisis y difusión de la información y elementos de autovaloración. La identificación de estas competencias informacionales constituyó un elemento clave por el cual las bibliotecas pueden orientarse para el desarrollo de acciones específicas en su comunidad.

**Palabras clave:** competencias informacionales, alfabetización informacional, encuesta

**Abstract**

The present article shows the results of a survey aimed at identifying the informational abilities of Havana University students. Several methods such as the survey, expert's interviews and content and document analysis are used. The questionnaire has been structured base on three basic variables: information search, information analysis and release and self-evaluation elements. The identification of these abilities was a key element for guiding libraries in the development of actions focused on their communities.

## Introducción

Los estudios sobre Alfabetización Informacional (ALFIN) son una temática de profundidad y protagonismo en las bibliotecas universitarias. Su finalidad es asegurar que los usuarios sepan cómo tienen que aprender, y por qué necesitan hacerlo con respecto a sus relaciones con las fuentes de información, y no solo las que están en la biblioteca, sino también con todas las del ámbito social, (Bawden, 2002). Se reconoce como prioridad básica que el estudiante de la enseñanza superior posea las competencias adecuadas en cuanto al uso y manejo de la información. Es por ello que la universidad juega un papel importante, al igual que el de sus profesionales de la información; cuyas funciones también van encaminadas a formar a sus comunidades de usuarios. García (2007) reconoce que «el universitario actual dispone de más información de la que puede procesar, por lo que una de las funciones de la enseñanza universitaria sería la de facilitar al alumno las herramientas (cognitivas y conceptuales) que le ayuden a procesar la información más relevante. La universidad debe intentar conseguir alumnos críticos, dotados de los conocimientos, habilidades y actitudes que le permitan seleccionar, procesar, analizar y sacar conclusiones de las informaciones que recibe y ser capaz de exponerlas a través de diferentes medios.»

En el contexto específico de la Biblioteca Central de la Universidad de La Habana, la Dirección de Información (DI), quien rectora los servicios informacionales de la comunidad universitaria, llevó a cabo en los dos últimos años (2010-2011) un rediseño de su línea de trabajo; evidenciándose directamente en la ampliación de servicios que pasaron de ser tradicionales a estar mediados por el uso de la tecnología. Esto ha conducido a los estudiantes ha ejecutar nuevas prácticas antes no desarrolladas. Es por eso que surge como preocupación identificar hasta qué punto éstos poseen las competencias apropiadas para interactuar con estos nuevos servicios y recursos. Dicha identificación de competencias se abordó desde una inclusión informacional; a partir de la cual se profundizó en los aspectos cognitivos de estos individuos como usuarios de la información (Marciales et. al., 2008). Paralelo a lo anteriormente enunciado se declara que el objetivo principal de este estudio es el de *presentar los resultados de una encuesta relacionada con la identificación de competencias informacionales que se les aplicó a los estudiantes de la Universidad de La Habana*. A través de sus resultados no solo se logrará que el estudiante autoevalúe su nivel de competencias, sino que serán develados los indicadores factibles con los cuales se podrán diseñar futuros entornos de enseñanza-aprendizaje (Gratch, 2006).

---

[1] Obtuvo la investigación el 1er lugar en el Fórum Científico-Estudiantl/2012 de la Facultad de comunicación de la Universidad de La Habana

## Metodología

Este artículo es un estudio exploratorio que tributa a la identificación de las competencias informacionales que los estudiantes de la Universidad de La Habana poseen. Las preguntas de interés se definen en tres variables esenciales: *búsqueda y recuperación de la información*, *análisis y difusión de la información* y *elementos autovalorativos;* las cuales agrupan algunas de las aptitudes que sobre el tema destaca Bernhard (2002). Se utilizaron como métodos la encuesta y como técnica de éste, el cuestionario (Véase anexo), también la entrevista no estructurada a expertos y por último el análisis documental y de contenido; tanto para abordar algunos aspectos teórico-conceptuales relacionados con la temática como para el análisis de los datos obtenidos en la encuesta. Una vez confeccionado el cuestionario se les envió a todos los estudiantes a través de la herramienta Lime Survey; la cual facilita la gestión de encuestas en línea, así como el procesamiento y análisis de los resultados. Se mantuvo la encuesta abierta durante un periodo de 10 días y tras haber culminado el plazo establecido se procesaron solo las respuestas completas que se obtuvieron. Para la representación gráfica de algunos datos se utilizó el programa Microsoft Excel.

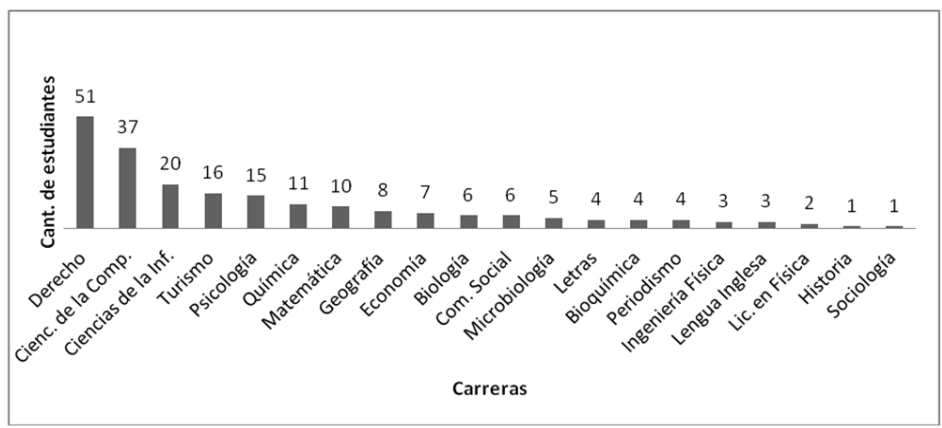

*Gráfico 1. Listado de carreras.*

## Resultados

Se obtuvo un total de 214 respuestas en las cuales se les preguntó en primera instancia la carrera a la cual pertenecían, respondiendo 20 de un total de 29 carreras. (Véase gráfico 1)[1]. Con respecto al año que cursaban alcanzaron tener un mayor porcentaje los estudiantes de segundo año (24.77%), sucediéndoles los de primero y cuarto año (ambos con un 20.09%) y luego los de tercero (19.16%) y quinto año (15.89%). El interés de respuesta disminuyó en aquellos que ya están culminando sus estudios.

## Búsqueda de información

A continuación se exponen los resultados asociados a la primera de las variables de este estudio. Respecto a la búsqueda de información el 80.37 % de los encuestados aseguró necesitar «siempre» de información para el estudio de clases y la investigación, mientras que el 19.63% aseveró de necesitarla «a veces». No hubo respuesta para los que «nunca» requerirían de información para desarrollar sus actividades curriculares. Esto da la medida del alto reconocimiento existente en el estudiantado hacia la información como un componente básico en su formación y desenvolvimiento como profesional. El factor principal que impulsa su proceso de búsqueda es el interés personal (71.96%), seguido de los profesores (71.5%) los exámenes (60.28%) y por último, «otros» factores (5.14%). Con estas respuestas se evidenció que la búsqueda es una competencia meramente personal a pesar de que existen múltiples factores que la estimulan.

El gráfico 2 muestra los resultados sobre el lugar donde los estudiantes ejecutan las búsquedas. Se evidencia una marcada transición de visitas a espacios tradicionales, como las Bibliotecas de las diversas facultades (34.11%) o la Biblioteca Central (23.36%), a los espacios virtuales, como la Internet (74.3%) y la Intranet (49.07%).

Dentro del aspecto *fuentes de información* (véase gráfico 3) se obtuvo como la más consultada a las conferencias (77.57%); reconociéndose como la que más aporta conocimientos. En un segundo plano, las respuestas apuntaron hacia los sitios web (76.74%), quienes definen la tendencia actual de autogestión de contenidos. A través de dicho gráfico se evidencia claramente los valores secuenciales de respuesta sobre las consultas, las cuales son más frecuentes en aquellos medios que son de más fácil localización en el ambiente donde los estudiantes se desenvuelven.

Fue necesario conocer si se ha recibido preparación alguna en cuanto al uso de herramientas tecnológicas como internet e intranet. Los resultados reflejados en el gráfico 4 demuestran que la preparación se ha dirigido más sobre la intranet, debido a que casi siempre (20.56%) y siempre (21.96%) se les prepara para trabajar con ella, en lugar de Internet (casi siempre, 17.35% y siempre, 15.42%). Esto refleja que no existe una linealidad en la formación de habilidades, ya que a los estudiantes no se les enseña a trabajar más con internet; siendo ésta, según

---

[1] En el gráfico 1 no se representan las carreas que no respondieron la encuesta. Estas fueron: Historia del Arte, Estudios Socioculturales, Contabilidad y Finanzas, Farmacia, Alimentos, Filosofía, Lengua Francesa, Lengua Alemana y Lengua Rusa.

los datos del gráfico 2, la que más se usa para la búsqueda de información.

Sobre la consulta a bases de datos se evidenció una ínfima cultura de acceso (Véase gráfico 5). El Google Académico es la más consultada debido a que es uno de los más prestigiosos servicios que brinda el popularizado motor de búsqueda *Google*. Se ofrecen respuestas a través de al encuesta sobre consultas a bases de datos a las cuales los estudiantes acceden por diversas vías y a las que la Universidad esta suscrita. Para profundizar en otros aspectos relativos a la búsqueda de información se les preguntó si conocían sobre la existencia de motores de búsquedas. Ante ello el 53.27% formuló que no, mientras que el 46.73% afirmó conocer al respecto. La encuesta destacaba una lista de estos motores para que fuesen seleccionados los que habitualmente se usaban. Entre ellos figuraron Google (46.26%), Yahoo Search (30.37%) y Microsoft (11.68%). Si se comparan estos datos con los de la respuesta anterior se sigue distinguiendo al Google y a sus servicios como la prioridad y la referencia suprema dentro del ambiente Web.

Un 63.08% considera redefinir «a veces» la búsqueda que realiza, mientras que el 29.44% alega «sí» redefinirla y un 7.48% «no» lo hace. Esto está influenciado por el uso de los términos relacionados con las materias afines que se dominan según la especialidad. Con respecto al uso de los operadores booleanos el 74.77% refiere no usarlo durante sus procesos de búsquedas. Esto pudiera estar dado por el desconocimiento sobre la existencia de los mismos y de las funcionalidades que posee.

Era importante conocer las diversas vías a través de las cuales los estudiantes han adquirido determinadas competencias, específicamente con respecto a la búsqueda de información en Internet. Se aseveró haberlas adquirido, en su mayoría, de manera autónoma (65.42%) y a través de los compañeros (18.69%); para ello véase gráfico 6.

Es decir, que han tenido gran protagonismo ciertos espacios de interacción y grupos afines que tal vez están fuera del área académico-universitaria, como parte de las evidentes políticas de accesibilidad, que comparte la sociedad. Se puede inferir que la adquisición de este tipo de competencia tiende a generarse a través de procesos poco formalizados.

Era importante conocer las diversas vías a través de las cuales los estudiantes han

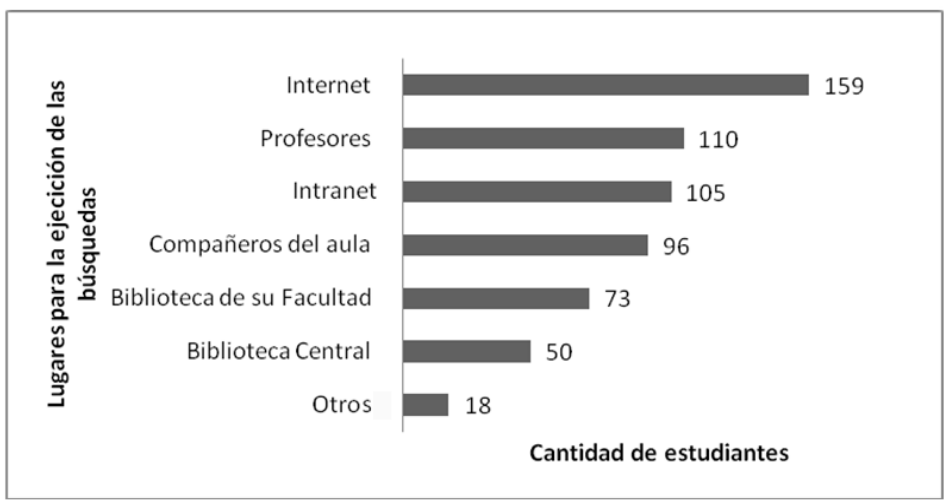

*Gráfico 2. Lugares frecuentados para la búsqueda de información.*

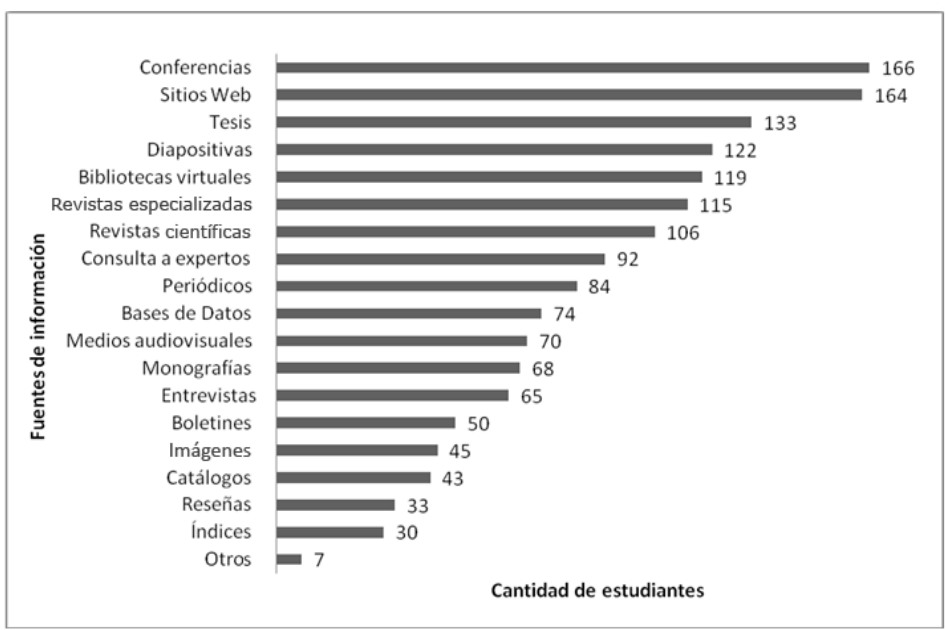

*Gráfico 3. Fuentes de información consultadas.*

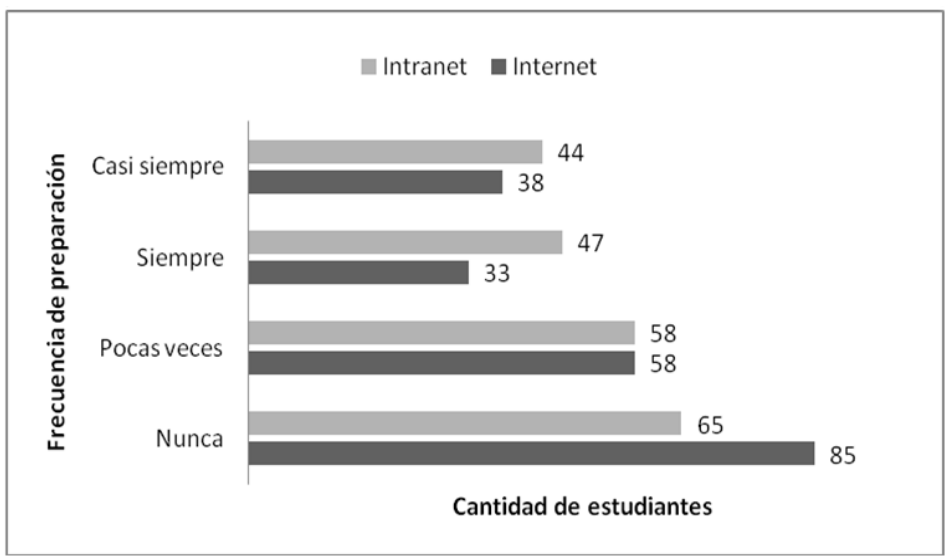

*Gráfico 4. Preparación en cuanto al uso de Internet e Intranet.*

adquirido determinadas competencias, específicamente con respecto a la búsqueda de información en Internet. Se aseveró haberlas adquirido, en su mayoría, de manera autónoma (65.42%) y a través de los compañeros (18.69%); para ello véase gráfico 6.

Es decir, que han tenido gran protagonismo ciertos espacios de interacción y grupos afines que tal vez están fuera del área académico-universitaria, como parte de las evidentes políticas de accesividad, que comparte la sociedad. Se puede inferir que la adquisición de este tipo de competencia tiende a generarse a través de procesos poco formalizados.

El uso de los servicios de la Biblioteca Central de la Universidad de La Habana, como rectora de la red de bibliotecas de toda la comunidad universitaria, fue otro de los aspectos a medir. Un 68.69% confirmó no conocer a plenitud de los servicios que se ofrecen. Los que respondieron positivamente, conocerlos todos, señalaron usar más los de la Sala de Servicios Digitales (29.78%), el Catálogo Electrónico (20.56%) y el Préstamo de Documentos (19.14%). En el gráfico 7 se muestra de una forma más ampliada los resultados más relevantes. Como se puede apreciar existe una tendencia de explotar mayormente los servicios digitales en lugar de los tradicionales. Otra recurso que vale destacar, son las revistas científicas, que aportan documentos evaluados de destacados investigadores, los cuales incluimos como recurso al consultar las bases de datos internacionales como Scopus, Latindex, Redalyc y otras.

## Análisis y difusión de la información

Asociado a esta variable fue de vital importancia identificar si los estudiantes usaban criterios para validar la calidad de la información. El 71.96% afirma no conocer al respecto, mientras que el 28.04% sí afirmó conocerlos. Los que respondieron afirmativamente señalaron como los de frecuente empleo la actualidad (27.1%), el prestigio autoral (26.17%) y la accesibilidad (15.42%). En el gráfico 8 se expone la secuencia de uso de estos criterios. Aquellos que poseen un mayor valor pueden estar dados por la referencia directa de la información o bibliografía afín a su especialidad, que desde la docencia les son facilitadas.

Un 66.82% de los estudiantes conoce y aplica, de cierta forma, el principio de socialización

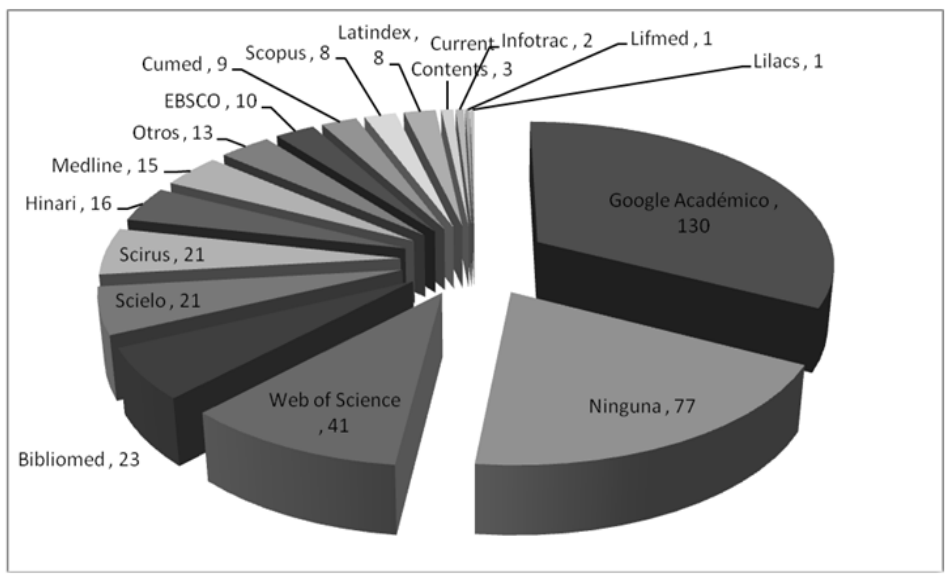

*Gráfico 5. Consulta de bases de datos.*

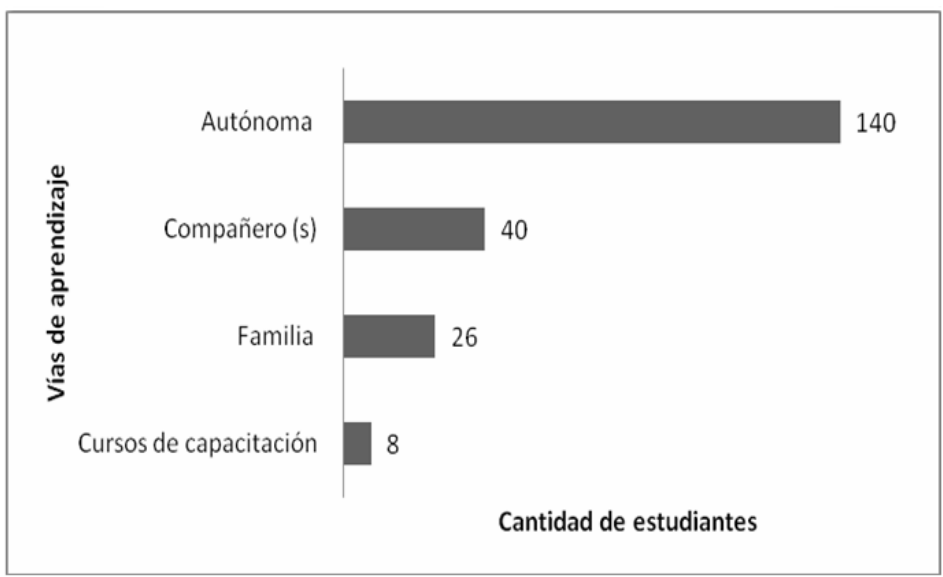

*Gráfico 6. Vías de aprendizaje para la búsqueda de información en Internet.*

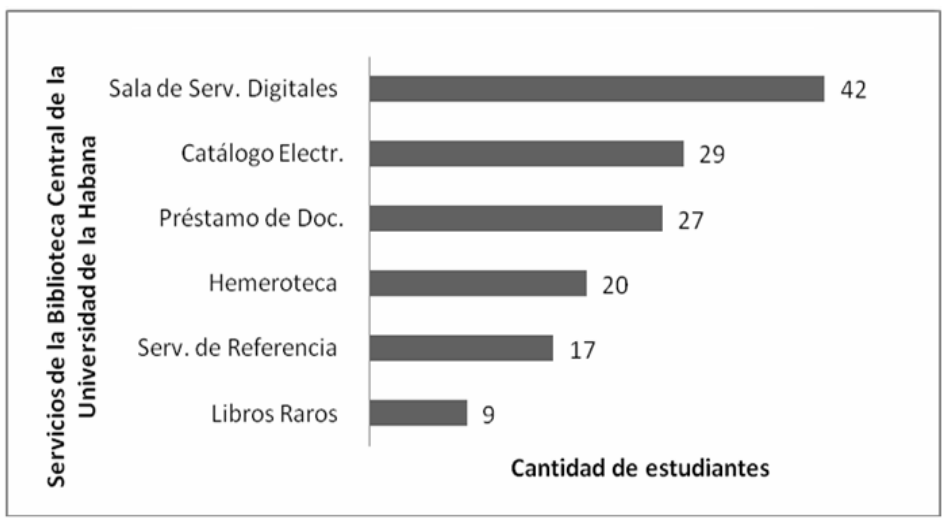

*Gráfico 7. Uso de los servicios de la Biblioteca Central de la Universidad de La Habana.*

de la información, mientras que el 32.71% atribuye que «a veces» lo hace y el 0.47 «nunca» lo hace. El conocimiento se socializa a partir de que la información se comparte; contribuyendo a que se establezcan amplias redes de difusión; es por eso que sobre la base de esta idea será necesario conocer los medios a través de los cuales se difunde la información. El mayor rango lo alcanzó el correo electrónico (86.92%), ya que éstos en su mayoría son propios de la institución universitaria y por lo tanto no generan problemas de accesibilidad. En escalas inferiores figuraron otros medios como los foros (17.29%) y chats (11.68%); por lo que también se presentan índices de difusión a través de internet, los cuales no deben ser obviados.

La elaboración de bibliografías a partir de los diversos estilos que existen es una habilidad que el estudiante de la enseñanza superior debe incorporar, tanto a las actividades curriculares y docentes, como a la investigación; conocen esto el 54.21% y cómo aplicar tales estilos, el 45.79% no. Los que respondieron afirmativamente declararon el uso de las normas APA (23.26%), Harvard (10.28%), ISO (9.35%) y Vancouver (7.01%). Los gestores bibliográficos, como herramientas técnicas que ayudan la elaboración de bibliografías, es del desconocimiento del 83.64% de los encuestados y dentro del pequeño grupo que los conocen se resaltó EndNote (16.62%) como el más manejado.

### Elementos de autovaloración

Una valoración autocrítica de los encuestados respecto a sus habilidades en la búsqueda de información reflejó que el 50.93% las considera «regular», mientras que el 43.46% las cree «buenas» y el 5.61% las clasifica como «malas». Esto da la medida de que el estudiante en muchas ocasiones no logra satisfacer sus necesidades informacionales y lo relaciona a sus malas competencias para desarrollar este tipo de actividad. El 98.13% atribuye de importante conocer sobre las herramientas de búsqueda, uso y manejo de información. Relativo a ello, el 57.48% incide en que sería importante y necesario implementar acciones para trabajar mejor con la información y con cada uno de sus recursos. De forma notable también se destacó que el 62.15% concibe como apropiado la incorporación de una asignatura al plan de estudios de su carrera, la cual les aporte los conocimientos pertinentes al tema.

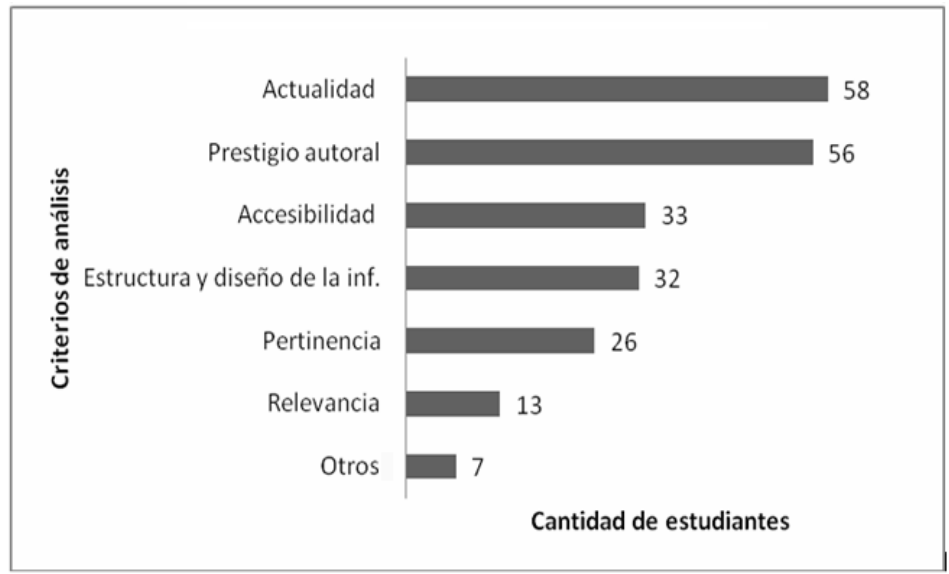

*Gráfico 8. Criterios para el análisis de la información.*

### Conclusiones

Dentro del proceso de formación académica de los estudiantes universitarios, el buen uso que se haga de la información contribuye a un mejor desenvolvimiento en sus actividades y tareas. Es prioridad de las bibliotecas universitarias medir el estado de las competencias informacionales que sus usuarios poseen. A partir de la aplicación del cuestionario se pudo determinar que los estudiantes tienden a hacer un elevado uso de servicios virtuales y reconocen su falta de habilidades para trabajar con la información que en esos ambientes se presenta. Consideran además que sería oportuno que se lleven a cabo acciones formadoras que les ayude a enfrentar y solucionar dichas deficiencias. La identificación de estas competencias informacionales constituye un elemento clave por el cual las bibliotecas y las direcciones docentes de las facultades deben orientarse para el desarrollo de acciones específicas en su comunidad.

### Agradecimientos:

### Referencias bibliográficas

Bawden, David. (2002). Revisión de los conceptos de alfabetización informacional y alfabetización digital [versión electrónica]. Anales de Documentación, 5. Disponible en: http://revistas.um.es/analesdoc/article/view/2261/2251.

Bernhard, Paulette. (2002). La formación en el uso de la información: Una ventaja en la enseñanza superior. Situación actual [versión electrónica]. Anales de Documentación, 5. Disponible en: http://revistas.um.es/analesdoc/article/view/2271/2261.

García Valcárcel, A. (2007). Herramientas tecnológicas para mejorar la docencia universitaria. Una reflexión desde la experiencia y la investigación [versión electrónica]. Revista Iberoamericana de Educación a Distancia, 10 (2). Disponible en: http://www.utpl.edu.ec/ried/images/pdfs/volumendiez/herramientas-tecnologicas.pdf.

Gómez-Hernández, J., Benito Morales, F., Cerdá Díaz, J., et al. Estrategias y modelos para enseñar a usar la información (2000). Disponible en: http://hdl.handle.net/10760/6717

Gómez Hernández, J. y Licea de Arenas, J. (2002) La alfabetización en

## Referencias

información en las universidades [Versión Electrónica]. Revista. de Investigación Educativa, 20 (2). Disponible en: http://www.doredin.mec.es/documentos/007200330098.pdf

Gratch Lindauer, B. (2006). Los tres ámbitos de evaluación de la alfabetización informacional [Versión Electrónica]. Anales de Documentación, 9. Disponible en: http://revistas.um.es/analesdoc/article/view/1411/1461.

Núñez Jover, Jorge (2003). La Ciencia y la Tecnología como procesos sociales. Editorial Félix Varela. La Habana.

Marciales Vivas, G., González Niño, L., Castañeda Peña, H. y Barbosa Chacón, J. (2008). Competencias informacionales en estudiantes universitarios: una reconceptualización [Versión Electrónica]. Revista Universitas Psychologica, 7(3) Disponible en: http://sparta.javeriana.edu.co/psicologia/publicaciones/actualizarrevista/archivos/V07N03A03.pdf.

**Est. 5to. Carlos Luis González Valiente**
Facultad de Comunicación.
Universidad de La Habana
País: Cuba
Correo electrónico: <carlos.valiente@fcom.uh.cu>

**Est. 5to. Yilianne Sánchez Rodríguez**
Facultad de Comunicación.
Universidad de La Habana
País: Cuba
Correo electrónico: <yilianne.sánchez@fcom.uh.cu>

**Est. 5to. Yazmín Lezcano Pérez**
Facultad de Comunicación.
Universidad de La Habana
País: Cuba
Correo electrónico: <yazmin.lezcano@fcom.uh.cu>

Anexo

**Identificación de competencias informacionales. Cuestionario para estudiantes de la Universidad de la Habana.**

Carrera: _______________________________________  Año: __________

**I. Búsqueda de información**

1. ¿Requiere de buscar, acceder y utilizar información de manera constante para la investigación y el estudio?
   Marque una opción. ___ Siempre  ___ A veces  ___ Nunca

2. ¿Qué incide con mayor frecuencia en su constante búsqueda de información para el estudio? Marque las opciones que correspondan.
   ___Profesores  ___Pruebas  ___Interés personal  ___Formación Profesional  ___Otros:_______________________

3. Para la búsqueda de información, ¿a dónde recurre con mayor frecuencia? Marque las opciones que correspondan.

4. ___Biblioteca de su Facultad  ___Internet  ___Profesores  ___Compañeros de aula  ___ Biblioteca Central
   ___Otro:_______________________

5. De las fuentes de información que se relacionan a continuación marque cuáles consulta con mayor frecuencia.
   ___ Monografías  ___Entrevistas  ___Consulta a expertos  ___Conferencias  ___Índices  ___Revistas impresas
   ___ Medios audiovisuales  ___Fotos  ___Sitios Web  ___Bibliotecas virtuales  ___Boletines electrónicos  ___Revistas electrónicas
   ___Bases de Datos  ___Diapositivas  ___Catálogos  ___Periódicos  ___Reseñas  ___Tesis  ___Otros.

6. ¿Te ofrecen en la facultad alguna preparación para trabajar con la información empleando las herramientas tecnológicas?
   · Información en Internet: __ Siempre  __Casi siempre  __Pocas veces  __Nunca
   · Información en Intranet: __ Siempre  __Casi siempre  __Pocas veces  __Nunca

7. De las siguientes bases de datos, ¿cuáles conoce o ha consultado alguna vez? Marque las opciones correspondientes.
   __ Bibliomed __Scielo __Scopus __Latindex __EBSCO __Cumed __Medline __Current Contents __Lifmed __Lilacs __Hinari
   __Infotrac __Web of Science __Google Académico __Scirus __Ninguna __Otra.

8. ¿Conoces qué son los motores de búsqueda?
   __Sí __No. Si es afirmativa su respuesta, marque cuáles ha utilizado: __Google __ Microsoft __ Kartoo __ Altavista __ Ixquic __ Terra
   __ Yahoo Search __ Ask __ Lycos __ Aol __ MSN __ Otro _____________

9. ¿Encuentra palabras o frases equivalentes para expresar la misma idea al solicitar información?
   Marque una opción. __Sí __No __A veces

10. ¿Utilizas los operadores booleanos (permiten enfocar la búsqueda vinculando palabras o frases: And/ Not/Or/XOR)?
    Marque una opción. __Sí ___No

11. ¿De qué manera ha aprendido a buscar información en Internet?
    Marque una opción. __ De manera autónoma __ A través de un compañero __ A través de cursos de capacitación
    __ Con miembros de su familia __Sin respuesta.

12. ¿Conoce los servicios de información que brinda la Biblioteca Central?
Marque una opción. __ Sí __ No. De ser afirmativa la respuesta anterior indique cuáles ha utilizado: __ Préstamo en sala de lectura __ Préstamo interbibliotecario __ Consulta online __ Servicio de Referencia __ Libros Raros ___ Hemeroteca.

**II. Análisis y difusión de la información.**

1. ¿Conoces los criterios que existen para validar la calidad de las fuentes que consultas?
   Marque una opción. __Sí __No

2. ¿De ellos cuáles ha tenido en cuenta?
   Marque las opciones que correspondan. __Actualidad de la Información  __La información satisface tu carencia de conocimiento
   __ Prestigio del autor  __Consideras válida la información porque estás de acuerdo con lo que plantea el autor  __Accesibilidad de la información (no solo que te resultó fácil hallar la información, sino que su nivel intelectual te permite entender el contenido que en ésta presenta).

3. ¿Con qué frecuencia compartes la información que obtienes para pruebas, seminarios y trabajos con tus compañeros?
   ___ Siempre ___ A veces ___ Nunca.

Anexo (continuación)

4. ¿Qué medios utilizas para la difusión de información?
   Marque las opciones que correspondan: __Correo electrónico  __Chat  __Blogs  __ Foros  __Listas de discusión  __Ninguno.

5. ¿Sabes elaborar bibliografías?
   Marque una opción.  __ Sí  __No

6. De los estilos bibliográficos existentes, ¿cuál utilizas?
   Marque una opción: __APA  __Vancouver  __ISO  __Chicago  __ Harvard  __ Ninguno  __Otros:

7. ¿Conoce los gestores bibliográficos?
   Marque una opción. __Sí __No. En caso de ser afirmativa su respuesta, ¿cuál ha utilizado?
   __EndNote __Zotero __ Procite __ Reference Manager __ Mendeley __Otros: __________________

**III. Elementos de autovaloración.**

1. ¿Cómo evalúas tus habilidades personales para la búsqueda de información relacionada a sus temas de investigación?
   Marque una opción.  ___ Buenas  ___ Regulares  ___ Malas

2. ¿Considera importante conocer las herramientas de búsqueda, uso y manejo de información en su especialidad?
   Marque una opción.  ___ Sí  ___ A veces  ___ No

3. ¿Cree que es factible implementar acciones para el desarrollo de competencias relacionadas con el uso y manejo de la información?
   ___ Sí  ___ No

4. ¿Considera necesario la incorporación de una asignatura que aporte los conocimientos relacionados al uso y manejo de la información?
   ___Si  __No  __Quizás.